\documentclass[12pt]{article}
\begin{document} 


\title{\bf QCD$_4$ Glueball Masses from AdS$_6$ Black Hole Description }
\author{{\sc Cong-Kao Wen,   \sc Huan-Xiong Yang \thanks{E-mail: hxyang@zimp.zju.edu.cn}}
\\
{~~}
\\
{\it Zhejiang Institute of Modern Physics, Physics Department,}
\\
{\it Zhejiang University, Hangzhou, 310027,\, P. R. China}
\\
{~~}
\\
{~~}
\\
{~~}}

\date{\today}
\maketitle

\begin{abstract}

By using the generalized version of gauge/gravity correspondence, we study the mass
spectra of several typical QCD$_4$ guleballs in the framework of AdS$_6$ black hole
metric of Einstein gravity theory. The obtained glueball mass spectra are numerically
in agreement with those from the AdS$_7 \times$S$^4$ black hole metric of the
11-dimensional supergravity.
\end{abstract}

\maketitle

\newpage
The idea that the non-perturbative aspects of 4-dimensional QCD may have a dual
description in terms of supergravity limit of string theories is perhaps among the most
appealing concepts in particle physics. With the advent of D-branes\cite{Pol}, the
AdS/CFT correspondence\cite{Maldacena} and then the thermal mechanism for breaking
supersymmetry and conformal symmetry\cite{Witten}, studies have been intensified for
phenomenologies of this expanded holographic principle in recent years. An approach to
QCD from the gauge/gravity correspondence is to estimate glueball masses of QCD (in the
large $N$ limit) from supergravity models in higher dimensions. Although the
supergravity approximation is not satisfactory for a serious study of large $N$ QCD, it
is surprising that the mass spectra of QCD glueballs obtained from the supergravity
dual descriptions have qualitatively coincided with those from lattice QCD calculations
\cite{Koch, Csaki, Brower, Braga, Constable}\cite{ Peardon}.

There have basically been two kinds of approaches for calculating QCD$_4$ glueball
masses from the conjectured gauge/gravity correspondence. The first one consists of
starting from the dilaton-free truncations of the 11-dimensional supergravity and
breaking supersymmetries with different compactifications for bosonic and fermionic
degrees of freedom\cite{Witten, Brower}. This involves at least one compact space-like
dimension (the so-called thermal circle $S^1$) where anti-periodic boundary conditions
are assumed for the fermionic fields while the bosonic fields remain periodic. It leads
to a $AdS_7 \times S^4$ AdS-Schwarzschild black hole metric that can be related to
QCD$_4$. The second one begins with a slice of the AdS$_5$ metric of 5-dimensional
Einstein gravity theory\cite{Braga}. The brane on which QCD$_4$ fields and interactions
are constrained is assumed to be a 4-dimensional boundary of this AdS$_5$ slice.

With the perspective of the first kind of approaches, it appears more natural to
conjecture that the dual description of the QCD$_4$ should be such a gravity theory
that is only involved six dimensions of spacetime. Besides having the 4-dimensional
spacetime on which the QCD$_4$ interactions lie, it has to have a thermal circle and an
extra dimension. Based on this understanding, in this paper we intend to calculate the
large $N$ QCD$_4$ glueball masses from the AdS$_6$ AdS-Schwarzschild black hole metric.
The AdS$_6$ AdS-Schwarzschild black hole metric is difficult to be identified as a
non-BPS brane solution of Type II supergravity, however, it is an exact solution of
6-dimensional Einstein gravity theory with non-vanishing cosmological
constant\cite{Constable, Petersen}. In view of the general holographic principle
between gauge theories and gravity\cite{Susskind}, at least in estimating the glueball
masses of QCD$_4$, we conjecture that AdS$_6$ AdS-Schwarzschild black hole metric
provides a probable dual description of QCD$_4$ gauge theory.

Let us begin with the near-horizon AdS$_6$ AdS-Schwarzschild black hole metric ,
\begin{equation}\label{eq: 1}
ds^2=(r^2-\frac{1}{r^3})d\tau^2+ (r^2-\frac{1}{r^3})^{-1}dr^2 +r^2\sum_{i=1}^{4}dx^2_i
.
\end{equation}
The 6-dimensional bulk is described by a radial coordinate $r$ and a $4$-dimensional
Euclidean space-time coordinates $x_i ~(i=1,~2,~3,~4)$, a radial coordinate $r$, and
the thermal coordinate $\tau$. The $4$-dimensional Euclidean space-time is assumed to
be an effective 3-brane on which the QCD$_4$ glueballs live (we regard $x_4$ as the
time coordinate). This metric defines a horizon at $r=1$ so that the physically
relevant region is within the region $1\leq r < \infty$. The thermal coordinate $\tau$
is necessary to be periodic in order to avoid a conical singularity at the horizon and
break conformal symmetries. In writing Eq.(\ref{eq: 1}) we have adopted the units
$R_{AdS}=1$ so that the period of $\tau$ is $\beta={4\pi}/5$\cite{Witten,Petersen}. If
the theory is embedded into some supergravity/superstring framework, we have to impose
further the anti-periodic boundary conditions for fermionic degrees of freedom to break
the probable Supersymmetries. The gravitational fluctuations $h_{MN}$ are defined by,
\begin{equation}\label{eq: 2}
\bar{g}_{MN}=g_{MN}+h_{MN}
\end{equation}
where $g_{MN}$ denotes the AdS$_6$ background metric (\ref{eq: 1}) which is a solution
of six dimensional Einstein field equation with a negative cosmological constant:
\begin{equation}\label{eq: 3}
R_{MN}=-5 g_{MN}
\end{equation}
By the generalized gauge/gravity conjecture, the gravitational fluctuations in AdS$_6$
bulk are nothing but the glueballs of the large $N$ QCD$_4$ gauge theory on the
4-dimensional boundary $x_i ~(i=1,\cdots,4)$. For simplicity these metric perturbations
have no dependence on the spatial coordinates $x_i (i=1, 2, 3)$ and compactified
thermal coordinate $\tau$. Because glueballs are free particles at $N$ $\rightarrow
\infty$, we make an ansatz $h_{MN}=H_{MN}(r)e^{-mx_4}$ where $H_{MN}(r)$ is the radial
profile tensor and $m$ is the mass of the corresponding QCD$_4$ glueball. It follows
from Einstein equation $\bar{R}_{MN}=-5 \bar{g}_{MN}$ that the metric perturbations
$h_{MN}$ satisfy the equations of motion:
\begin{equation}\label{eq: 4}
\frac{1}{2}\nabla_M \nabla_N h^L_L +\frac{1}{2}\nabla^2 h_{MN} - \nabla^L
\nabla_{(M}h_{N)L}-5h_{MN}=0~.
\end{equation}
They will be solved as an eigenvalue problem for determining the mass $m$ of QCD$_4$
glueballs.

As a tensor in 6-dimensional bulk the metric perturbation $h_{MN}$ has a variety of
polarizations (with total number 21). However, not all of these polarizations are
independent. Relying on the diffeomorphism invariance of the gravity theory under the
infinitesimal transformation of the coordinates $x^{M} \rightarrow
{x^{\prime}}^{M}=x^{M}+\epsilon^{M}(x)$, the equations of motion $(\ref{eq: 4})$ of the
perturbations are symmetric under the following ``gauge transformations'':
\begin{equation}\label{eq: 5}
{h^{\prime}}_{MN}(x)=h_{MN}(x)-g_{ML}(x)\partial_{N}\epsilon^{L}(x)
-g_{NL}(x)\partial_{M}\epsilon^{L}(x)-\epsilon^{L}(x)\partial_{L}g_{MN}(x)~.
\end{equation}
Consequently, there are 12 components non-independent in the metric perturbation
tensor. Due to a manifest $SO(3)$ rotational symmetry in the hypersurface
$x_{i}~(i=1,2,3)$ of the background (See Eq.(\ref{eq: 1})), among the remaining 9
components of independent polarizations there are $5$ perturbations forming the Spin-2
representation of algebra $SO(3)$, $3$ perturbations forming the Spin-1 representation
and $1$ perturbation forming the Spin-0 representation. They are respectively dual to
tensorial, vectorial and scalar-like glueballs in QCD$_4$.

A half of the non-independent components of $h_{MN}=H_{MN}(r)e^{-mx_4}$ can be removed
by gauge fixing. We first
assume the ``transverse gauge'':
\begin{equation}\label{eq: 6}
H_{M4}(r)=0~,~~~~\forall M~.
\end{equation}
The other components are expressed as,
\begin{equation}\label{eq: 7}
\left. \begin{array}{lll} H_{\tau\tau}(r)=S_{1}(r)~, &H_{\tau
r}(r)=A(r)~, &H_{\tau i}(r)= \sqrt{r^4 - \frac{1}{r}}V_{i}(r)~, \\
H_{rr}(r)=S_{2}(r)~, &H_{ri}(r)=B_{i}(r)~, &H_{11}(r)=[r^2 T_{1}(r)+S_{3}(r)]~,\\
H_{12}(r)=r^2 T_{3}(r)~,
&H_{13}(r)=r^2 T_{4}(r)~, &H_{22}(r)=[r^2 T_{2}(r)+S_{3}(r)]~,\\
H_{23}(r)=r^2 T_{5}(r)~, & &H_{33}(r)=[S_{3}(r)- r^2 T_{1}(r)- r^2 T_{2}(r)]~.  ~~~~
\end{array}
\right.
\end{equation}
Substitution of these ansatz into Eq.(\ref{eq: 5}) leads to $A(r)=B_{i}(r)~(i=1,2,3)=0$.
Besides,
\begin{equation}\label{eq: 8}
\left. \begin{array}{l} r(r^5-1)T^{\prime \prime}_{i}(r)+(6r^5 -1)T^{\prime}_{i}(r)
+ m^2 r^2  T_{i}(r)=0~,\\
r(r^5-1)V^{\prime \prime}_{j}(r)+(6r^5 -1)V^{\prime}_{j}(r) +[m^2 r^2  -{25}/{4 r (r^5-1)}]V_{j}(r)=0~.
\end{array} \right.
\end{equation}
for $(i=1,2,3, 4, 5)$ and $(j=1,2,3)$ and where
$T^{\prime}_{i}(r)=\partial_{r}T_{i}(r)$, etc. From the perspective of the boundary
QCD$_4$, the former five independent perturbations $T_{i}(r)~(i=1,2,\cdots, 5)$ form
the spin-2 representation of algebra $SO(3)$ and the latter three perturbations
$V_{j}(r)~(j=1,2,3)$ form the spin-1 representation. These perturbations do
respectively represent a QCD$_4$ tensor glueball and a vector glueball. The remaining
perturbations are expected to form the spin-0 representation of $SO(3)$ so that a dual
description of the QCD$_4$ scalar glueball arises. As expected, only $S_3(r)$ in the
remaining fluctuations $S_i(r)~(i=1,~2,~3)$ is independent, which satisfies a third
order differential equation:
\begin{equation}\label{eq: 9}
\left. \begin{array}{l} r^3 (r^5 -1)^2 [4 (r^5 -1) + m^2 r^3] S^{\prime
\prime\prime}_3(r) + r^2 (r^5 -1) [8 (r^5 -1) (2r^5 +3) \\
~~~~+ m^2 r^3 (2r^5 -m^2r^3 +13)] S^{\prime \prime}_3(r) - r [16 (r^5 +4 )(r^5 -1)^2
\\~~~~ + m^2 r^3 (4 r^{10} + m^2 r^8 -3 r^5 + 4 m^2r^3 -26) ] S^{\prime}_3(r) + [80 (r^5 -1)^2 \\
~~~~+ m^2 r^3 (4 r^{10} + 2m^2r^8 - m^4 r^6 -28 r^5 -2m^2 r^3 -26)] S_3(r)~=0~.
\end{array}
\right.
\end{equation}
Once $S_3(r)$ is known, the other two perturbations can be determined by the following
equations:
\begin{equation}\label{eq: 10}
\left. \begin{array}{l} 4 r^2(2 r^5 +3) (r^5 -1)^2 S^{\prime \prime}_3(r) + r (r^5 -1)[
4 (2 r^5 +3)^2 -15 m^2 r^3] S^{\prime}_3(r) \\ ~~~-2 (r^5 -1) [4 (3 r^5 +2) (2r^5 +3) +
m^2 r^3 (8 r^5 + 3 m^2 r^3 -3 )] S_3(r) \\ ~~~ -2 m^2 r^8 [4 (r^5 -1) + m^2 r^3] S_1(r)
~=~0~,
\end{array}
\right.
\end{equation}
and
\begin{equation}\label{eq: 11}
\left. \begin{array}{l} 15 r^2 (r^5 -1) S^{\prime}_3(r) + 6 r (r^5 -1) (m^2 r^3 -5)
S_3(r) \\ ~~~-2 r^6 (10 - m^2 r^3) S_1(r) - 4 (2 r^5 +3) (r^5 -1)^2 S_2(r)~=~0~.
\end{array}
\right. \end{equation}

It is difficult in general to solve a differential equation of the order higher than
two. However, in the considered case, the ``gauge symmetry'' of Eq.(\ref{eq: 4}) enable
us to reduce the order of this equation for determining the QCD$_4$ scalar glueball
masses to two. Our strategy is to find another gauge in which $H_{44}(r)$ does still
remain vanishing. Choosing the (nonzero) gauge transformation parameters as follows:
\begin{equation}\label{eq: 12}
\epsilon ^{r}(r)=S_3(r)/{2r}~,~~~~\epsilon^{4}(r)=S_3(r)/{2mr^2}~,
\end{equation}
we get from ansatz (\ref{eq: 7}) the expressions of the metric perturbations in the new
gauge:
\begin{equation}\label{eq: 13}
\left. \begin{array}{l} H^{\prime}_{\tau\tau}(r)=S_{1}(r) - (1 + 3/{2r^5})S_3(r)~,\\
H^{\prime}_{rr}(r)=S_{2}(r)+ r[(4r^5 +1)S_3(r)-2r(r^5 -1)S^{\prime}_3(r)]/{2(r^5 -1)^2}~,\\
H^{\prime}_{r4}(r)=-S^{\prime}_3(r)/{2m}~ +[2(r^5 -1)+ m^2 r^3]S_3(3)/{2mr(r^5 -1)} ~,\\
H^{\prime}_{11}(r)= r^2 T_{1}(r)~,\\
H^{\prime}_{22}(r)= r^2 T_{2}(r)~,\\
H^{\prime}_{33}(r)= - r^2 T_{1}(r)- r^2 T_{2}(r)~.  ~~~~
\end{array}
\right.
\end{equation}
The other perturbations have the same expressions as those in the transverse gauge.
These perturbations in the new gauge are in fact some special combinations of the old
ones and their derivatives (with respect to coordinate $r$) in transverse gauge. We
expect that they may satisfy the conventional two order equations. By a tedious but
straightforward calculation, we find that $H^{\prime}_{\tau \tau}(r)$ is indeed subject
to a two order differential equation:
\begin{equation}\label{eq: 14}
r(r^5 -1)\frac{d^2}{dr^2}S(r) +(6r^5 -1)\frac{d}{dr}S(r) +[m^2 r^2 + \frac{750
r^4}{(8r^5 -3)^2}]S(r)=0~
\end{equation}
where,
\begin{equation}\label{eq: 15}
S(r):={r^3 H^{\prime}_{\tau \tau}(r)}/{(8 r^5 -3)}={r^3 [S_1(r)-(1+
3/{2r^5})S_3(r)]}/{(8 r^5 -3)}~.
\end{equation}

We now to calculate the discrete mass spectrum for these three kinds of glueballs. It
follows from Eqs.(\ref{eq: 8}) and (\ref{eq: 14}) that the asymptotic behavior of their
linear independent solutions at $r \rightarrow \infty$ are
\begin{equation}\label{eq: 16}
T_i(r),~V_j(r),~S(r)\sim r^{-5},~~1~;~~~~\forall i,~j.
\end{equation}
and at $r=1$ are
\begin{equation}\label{eq: 17}
\left \{
\begin{array}{ll}
T_i(r),~S(r) \sim 1,~~\log(r-1)~,~~&~\\
V_j(r) \sim \sqrt{r-1},~~\sqrt{r-1}\log(r-1)~;~~& \forall i,~j.
\end{array} \right.
\end{equation}
In all cases the reasonable boundary conditions at $r=1$ are the ones without the
logarithmic singularity. At $r \rightarrow \infty$ the singular asymptotic behavior is
necessary for having a normalizable eigenstate. The expected boundary conditions which
are also compatible with the field equations read
\begin{equation}\label{eq: 18}
\left \{
\begin{array}{ll}
T_i(r_1)=1,~&~T^{\prime}_i(r_1)=-{m^2}/5~;\\
V_j(r_1)=0,~&~V^{\prime}_j(r_1)\rightarrow \infty ~; \\
S(r_1)=1,~&~S^{\prime}(r_1)=-6-{m^2}/5~.
\end{array} \right.
\end{equation}
at $r_1=1$ and
\begin{equation}\label{eq: 19}
\lim_{r \rightarrow \infty}T_i(r)=\lim_{r \rightarrow \infty}V_j(r)=\lim_{r \rightarrow
\infty}S(r)=0,~\forall i,~j.
\end{equation}
at $r_{\infty} \simeq \infty$. Generically there are no solutions to Eqs.(\ref{eq: 8})
and (\ref{eq: 15}) that not only satisfy these boundary conditions but can also be
represented as series expansions convergent throughout the whole physical region $1\leq
r < \infty$. In fact, the series expansions of solutions
\begin{equation}\label{eq: 20}
\left. \begin{array}{ll} T^{(\infty)}_i(r)=\frac{1}{r^5} + \sum\limits^{\infty}_{n=1}
a^{(\infty)}_{in} \frac{1}{r^{n+5}}~,~~ & V^{(\infty)}_j(r)= \frac{1}{r^5} +
\sum\limits^{\infty}_{n=1}
b^{(\infty)}_{jn} \frac{1}{r^{n+5}}~, \\
S^{(\infty)}(r)=\frac{1}{r^5} + \sum\limits^{\infty}_{n=1} c^{(\infty)}_{n}
\frac{1}{r^{n+5}}~,
\end{array} \right.
\end{equation}
do converge in the region $I(\infty)=\{ r|~1<r<\infty \}$, while the series expansions
\begin{equation}\label{eq: 21}
\left. \begin{array}{ll} T^{(1)}_i(r)=1 + \sum\limits^{\infty}_{n=1} a^{(1)}_{in}
(r-1)^n ~,~~ & V^{(1)}_j(r)=(r-1)^{\frac{1}{2}} + \sum\limits^{\infty}_{n=1}
b^{(1)}_{jn} (r-1)^{n+\frac{1}{2}} ~, \\
S^{(1)}(r)=1 + \sum\limits^{\infty}_{n=1} c^{(1)}_{n} (r-1)^n~,
\end{array} \right.
\end{equation}
are convergent only in the neighborhoods of the horizon $I_{T}(1)=\{ r|~0\leq (r-1) <1
\}$, $I_{V}(1)=\{ r|~0\leq (r-1) <1 \}$ and $I_{S}(1)=\{ r|~0\leq (r-1)
<1-\sqrt[5]{\frac{3}{8}}\simeq 0.177 \}$. The expansion coefficients are associated
with the squared masses of glueballs. Therefore, coincidence of these solutions in the
overlap, see $I(\infty)\cap I_{T}(1)$, yields a discrete set of eigenvalues $m^2_{n}$,
where $n$ is the number of zeros of the following vanishing Wronski determinant:
\begin{equation}\label{eq: 22}
\left |
\begin{array}{ll}
T^{(\infty)}_{i}(r_0) ~& ~ (T^{(\infty)}_{i})^{\prime}(r_0)\\
T^{(1)}_{i}(r_0) ~& ~ (T^{(1)}_{i})^{\prime}(r_0)
\end{array}
\right |=0~,~~
\end{equation}
for an arbitrary point $r_0 \in I(\infty)\cap I_{T}(1)$. In numerical calculations this
$r_0$ should be chosen far away from the endpoints of the regions $I_{T}(1)$,
$I_{V}(1)$ and $I_{S}(1)$ because we have to truncate the series expansions of
(\ref{eq: 20}) and (\ref{eq: 21}) into some polynomials\cite{Zyskin}. As a check to the
above calculation, we have also solved these boundary problems by an independent
Runge-Kutta methods in which the boundary conditions (\ref{eq: 18}) and (\ref{eq: 19})
play crucial roles. It turns out that the calculation results from these two schemes
coincide. The obtained mass spectrum of the QCD$_4$ glueballs (in units $R_{AdS}=1$)
are presented in {\sl Table 1}.

\vspace{3mm}
\begin{table}
\begin{center}
\begin{tabular}{|c|c|c|c|}
\hline \multicolumn{1}{|c|}{\textsf{~~~}} &
\multicolumn{1}{|c|}{{\textbf{S}~}(~$0^{++}$)} &
\multicolumn{1}{|c|}{{\textbf{V}~}(~$1^{-+}$)} &
\multicolumn{1}{|c|}{{\textbf{T}~}(~$2^{++}$)}\\
\hline n=0 & 2.52 & 5.00 & 4.06  ~\\
\hline n=1 & 6.20 & 7.73 & 6.69 ~ \\
\hline n=2 & 8.93 & 10.34 & 9.25 ~ \\
\hline n=3 & 11.54 & 12.91 & 11.79 ~\\
\hline n=4 & 14.11 & 15.45 & 14.31 ~\\
\hline
\end{tabular}
\end{center}
\caption{\textsf{QCD$_4$ Glueball Mass Spectrum $m_{n}$ from AdS$_6$-BH Approach}}
\end{table}
\vspace{3mm}

The spectrum of these glueball masses are quantitative in agreement with those from the
AdS$_7$-BH dual theory\cite{Brower} which is presented in {\sl Table 2}. In particular,
both approaches indicate that the mass gap obeys an inequality
$m(0^{++})<m(2^{++})<m(1^{-+})$, and the lowest mass scalar comes from the
gravitational multiplet. By consulting with the available lattice spectrum for pure
SU(3) QCD$_4$ \cite{Peardon}, the qualitative agreement is also good. Here we have to
emphasize that it is such a qualitative agreement that is important. Because the
supergravity approximation to the gauge/gravity duality is in a ``wrong
phase''\cite{Witten, Petersen}, a reasonable estimation based on it should not be
expected to give reliable quantitative results. Consequently, the conjectured duality
between QCD$_4$ and AdS$_6$ supergravity works well in estimating the mass gap of
nonperturbative QCD$_4$.

\vspace{3mm}

\begin{center}
\begin{table}
\begin{center}
\begin{tabular}{|c|c|c|c|}
\hline \multicolumn{1}{|c|}{\textbf{~~~}} &
\multicolumn{1}{|c|}{{\textbf{S}~}(~$0^{++}$)} &
\multicolumn{1}{|c|}{{\textbf{V}~}(~$1^{-+}$)} &
\multicolumn{1}{|c|}{{\textbf{T}~}(~$2^{++}$)}\\
\hline n=0 & 7.308 & 31.985 & 22.097 \\
\hline n=1 & 46.986& 72.489 & 55.584  \\
\hline n=2 & 94.485 & 126.174 & 102.456  \\
\hline n=3 & 154.981 & 193.287 & 162.722  \\
\hline n=4 & 228.777 & 273.575 & 236.400  \\
\hline
\end{tabular}
\end{center}
\caption{\textsf{QCD$_4$ Glueball Mass-squared $m^2_{n}$ from AdS$_7$-BH Approach}}
\end{table}
\end{center}

\vspace{3mm}

In conclusion, we have seen that the AdS$_6$ black hole metric of a 6-dimensional
Einstein gravitational theory can also be applied to estimate QCD$_4$ glueball mass
spectrum, even though it is difficult to be identified as a nonextremal bane solution
of the Type II supergravity. It would be very important to understand whether this
agreement is purely a numerical coincidence or whether there is a deeper mechanism
behind it. It is also important to know whether this conjectured AdS$_6$ black hole
dual description can be used to describe other non-perturbative properties of QCD$_4$.

\subsection*{Acknowledgments}
We would like to thank M.X.Luo and H.Y.Jin for valuable discussions. The work is
supported in part, by CNSF-10375052, the Startup Foundation of the Zhejiang Education
Bureau and CNSF-90303003. H.X. Yang has also benefitted from the support of Pao's
Foundation.

\end{document}